\documentclass[dvips]{article}
\usepackage{icrctc07}

\title{Design Study of a Low Energy IACT Array for Ground-Based $\gamma$-Ray Astronomy}
\shorttitle{Future Low Energy IACT Array}
\authors{A.~Konopelko, J.P.~Finley, G.~Urbanski}
\shortauthors{A. Konopelko and et al}
\afiliations{Purdue University, Department of Physics, 525 Northwestern Avenue, 
West Lafayette, IN 47907-2036}
\email{akonopel@purdue.edu}

\abstract{Recently, ground-based very high-energy (VHE) $\gamma$-ray astronomy 
achieved a remarkable advancement in the development of the observational 
technique for the registration and study of $\gamma$-ray emission above 
100 GeV. Construction of telescopes of substantially larger sizes than the currently used 12~m 
class telescopes can drastically improve the sensitivity of ground-based detectors 
to $\gamma$ rays of energy from 10 GeV to 100 GeV. Based on Monte Carlo 
simulations we have studied the response of an array of three large area imaging 
atmospheric Cherenkov telescopes (IACT) as a prototype for a future large-scale low energy 
ground-based experiment. The sensitivity of a three-telescope array as a function of optical reflector 
size was investigated here in detail.
}

\begin{document}
\maketitle

\section{Introduction}
\vspace*{-2mm}

The sourthern hemisphere HESS array of four 12~m IACTs, located in Namibia, 
and the northern hemisphere VERITAS array of four similary designed telescopes, 
located in Arizona, have proven many outstanding advantages of 
stereoscopic observations of VHE $\gamma$ rays with the ground-based 
detectors. The construction and commissioning of the MAGIC 
experiment, consisting of a single
17~m telescope, equipped with a fast response, high-resolution imaging camera,
and located at La Palma in the Canary Islands, 
has also been completed recently. 
In the first years of its operation, the MAGIC telescope 
has demonstrated its performance, comparable to that of HESS and VERITAS, in 
observations of $\gamma$ rays with energies above 100~GeV in addition to 
the unique capability of detecting $\gamma$-ray showers ranging in  
energy well below 100~GeV, though at a relatively low sensitivity level. 
Currently, the MAGIC collaboration is constructing a second 17~m telescope on 
the same site in order to enable stereoscopic observations, which can drastically 
boost the sensitivity of future detections of $\gamma$-ray fluxes in the 
sub-100~GeV energy domain. Motivated by a growing scientific interest of 
the astrophysical community in $\gamma$-ray observations at low energies, 
the HESS collaboration recently began construction of a single 28~m imaging 
telescope in the center of their current array. This telescope will be able to detect $\gamma$ 
rays with energy as low as 40~GeV, enabling a rather broad 
overlap in dynamic energy range of this instrument and a future advanced 
space-born experiment -- GLAST.

At present, there are two major competing views on the direction of instrument development  
for ground-based VHE $\gamma$-ray astronomy. 
A prodigious scientific study, performed recently with 
the HESS array of IACTs, has strongly motivated various ongoing considerations 
of substantial expanding the current, rather modest multi-telescope system to 
much larger arrays, composed of roughly 50 telescopes of 12~m aperture each \cite{ref4,ref5}. Such arrays would 
increase   
tremendously the detection area of $\gamma$ rays, and consequently raise 
the sensitivity with respect to the $\gamma$-ray fluxes above 100~GeV by at least a 
factor of ten. However, this approach may not necessarily achieve a significant 
improvement in angular resolution or a substantial reduction in the energy threshold.
These are now considered to be the major limiting factors in understanding 
the morphology 
and energy spectra of many well-established VHE $\gamma$-ray sources. 
In another approach, 
an array of a few telescopes of substantially larger aperture, e.g. 20-30~m , can drastically 
improve the sensitivity of ground-based $\gamma$-ray detectors, primarily in the sub-100~GeV 
energy range \cite{ref3}, while at the same time providing high-quality $\gamma$-ray 
observations at energies above 100~GeV.
Finally, a combination of 
both approaches could meet most of the scientific requirements determined for the 
next generation of ground-based instruments \cite{ref6}.
 
In this paper, 
based on the Monte Carlo simulations, we investigate the
sensitivity of a three-telescope system, considered as a prototype 
of future low-energy arrays, as a function of telescope aperture.
 
 \vspace*{-4mm}
 \section{Simulations}
 \vspace*{-2mm}
 
The atmospheric showers produced by $\gamma$ rays and protons 
were simulated using the numerical code described in \cite{ref1}. The 
primary energy of simulated showers was randomized within the energy 
range from 10~GeV to 10~TeV. Events were weighted according to a power-law 
primary energy spectrum. The maximum impact distance of the shower axis to
the centre of the telescope array was $10^3$~m. The detailed procedure of simulating  
the camera response accounts for all efficiencies involved in the process of Cherenkov
light propagation and acquisition. The overall efficiency of the photon-to-photoelectron
(ph.-e.)
conversion was $\sim$0.1. The standard `picture' and `boundary' technique was applied for 
image cleaning. The simulated images were parameterized using the standard 
measures of their angular extent and orientation in the telescope focal plane. Further 
details on the simulation procedure can be found in \cite{ref3}.

Using the same sample of simulated $\gamma$-ray and cosmic-ray showers, 
the response of the array of three identical telescopes of various aperture sizes, 
i.e. 17, 20, 24, and 30 m, was calculated. The telescope coordinates were set up according 
to the center-symmetric triangular layout. The spacial separation between the telescopes and 
the center of the array was 50~m. The imaging cameras consisted of 1951 photo-multiplier tubes 
of $0.07^\circ$ each. This results in a $3^\circ$ diameter field of view.    

\vspace*{-4mm}
 \section{Detection Areas and Rates}
 \vspace*{-2mm}
 
The standard triggering scheme of imaging cameras requires a 
signal in two or three adjacent pixels (PMTs) to exceed, simultaneously, 
a certain threshold measured in the number of photoelectrons. The choice of trigger threshold is 
usually constrained by a minimal allowed accidental trigger rate, which is induced by continuous 
illumination of the camera pixels by the night sky background light. This accidental 
trigger rate has to be substantially suppressed with respect to the actual rate of recorded air 
shower events. The individual pixel load due to reflected background light is 
proportional to the aperture of the telescope reflector. Evidently, a larger telescope 
requires a higher trigger threshold in order to suppress accidental trigger rate. 
Exploiting a two-fold coincidence trigger for 
telescopes with aperture sizes in the range from 17 to 28~m, the corresponding trigger 
threshold is expected to be about 9-15 photoelectrons, respectively. The simulated events were
accepted for further consideration if the images in at least two telescopes out of three 
passed the trigger criterion.

\begin{figure}[!t]
\centering
\includegraphics [width=0.44\textwidth]{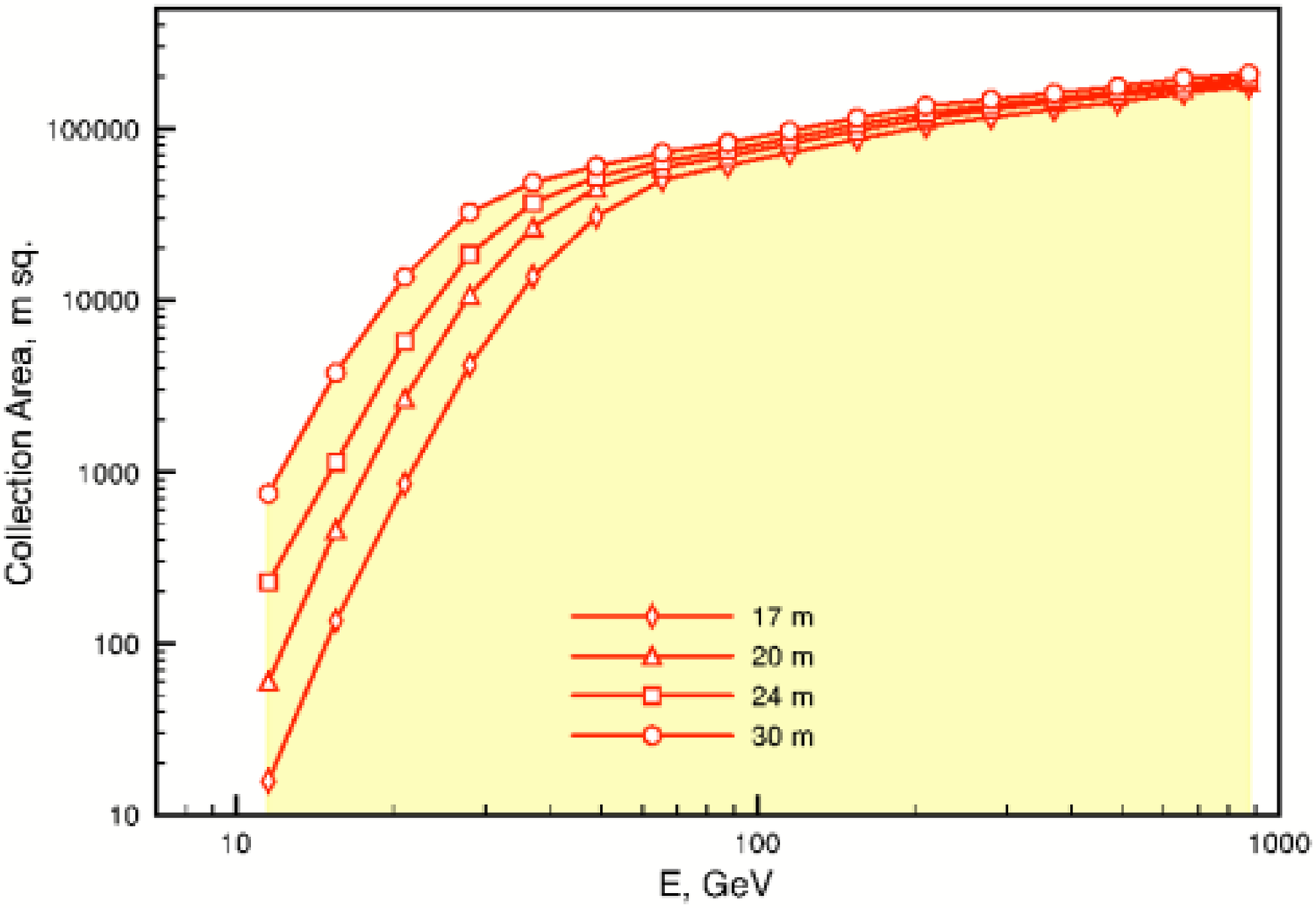}
\includegraphics [width=0.44\textwidth]{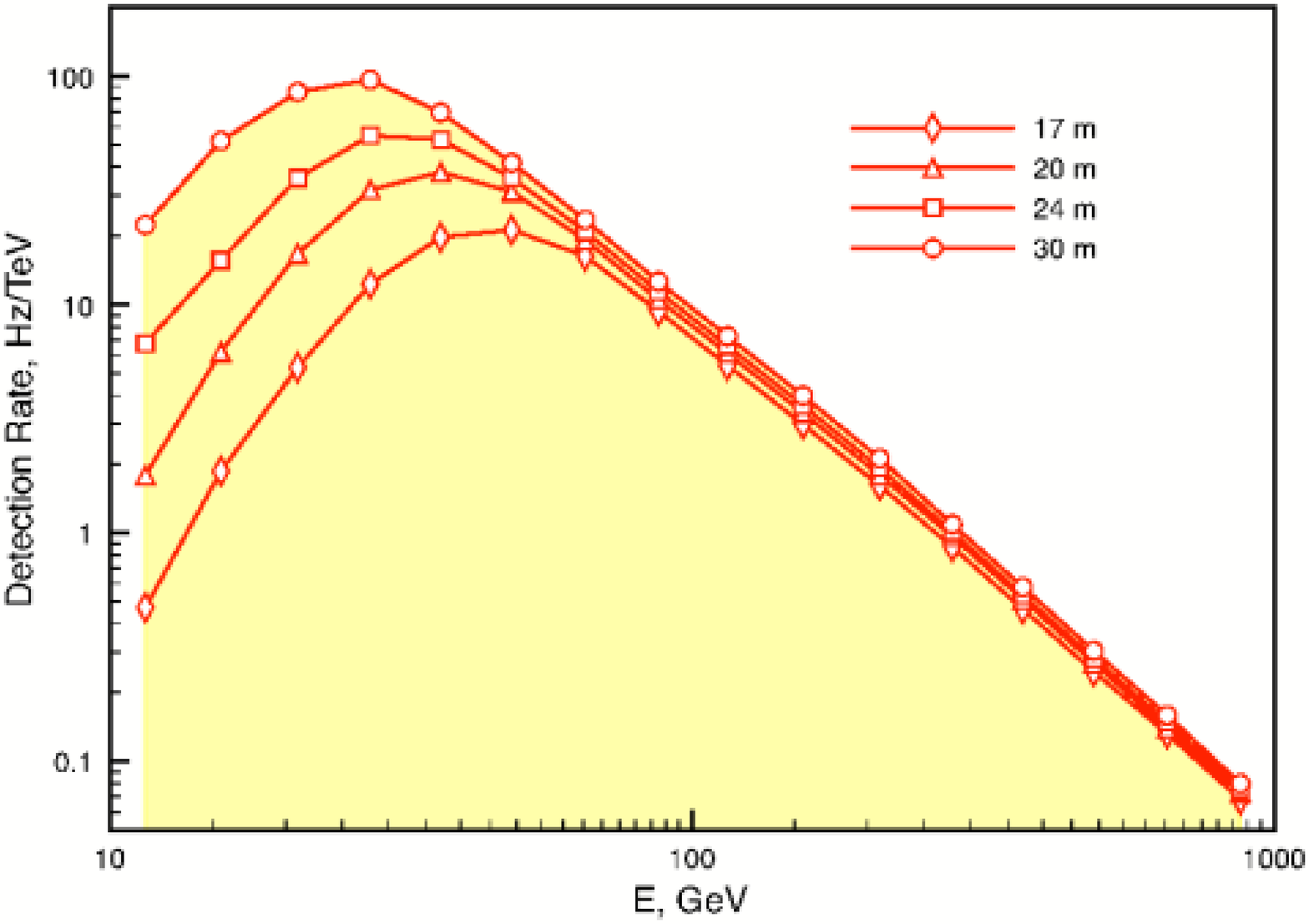}
\vspace*{-3mm}
\caption{Detection areas and rates of $\gamma$-ray showers for the array of three imaging 
telescopes of various aperture sizes.}
\label{fig1}
\vspace*{-3mm}
\end{figure}

Triggered $\gamma$-ray showers show a substantial increase in the detection area 
over the energy range from 10 to 100~GeV for telescopes of larger aperture (see Fig.~\ref{fig1}). 
Large telescopes are able to detect $\gamma$-ray showers of low energy at significantly 
larger impact distances. At the same time, in the energy range well above 100~GeV, telescopes 
of very different apertures yield almost identical detection areas. For these high energy $\gamma$-ray 
showers a rather narrow imaging camera of $3^\circ$ diameter is a major limiting factor. The images of the 
high-energy $\gamma$-ray showers, which are often registered at very large impact distances with 
respect to the telescopes, are focused onto the focal plane outside of the actual camera edge. 
 
\begin{figure}[t]
\centering
\includegraphics [width=0.41\textwidth]{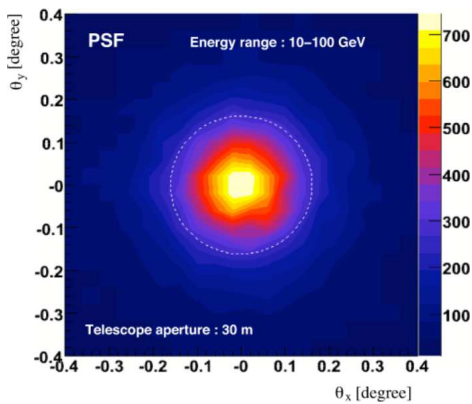}
\vspace*{-3mm}
\caption{The simulated point spread function (PSF) for a point-like $\gamma$-ray source observed with 
the array of three 30~m telescopes. The radius of the dashed white circle corresponds to the 
$\sigma$-parameter of a two-dimensional Gaussian fit. 
}
\label{fig2}
\vspace*{-3mm}
\end{figure}

Assuming a power-law energy spectrum of $\gamma$ rays as $dF_\gamma \propto E^{-2.6}dE$, 
which is close to the energy spectrum of the Crab Nebula (standard candle), 
one can compute corresponding differential detection rates (see Fig.~\ref{fig1}). It is apparent 
that the telescopes of larger apertures secure significantly lower energy threshold, 
defined as the peak 
energy of the differential detection rate. Increase in the telescope aperture from 17 to 30~m leads 
to a reduction of the energy threshold from 80~GeV down to 30~GeV. It is worth noting that the array of 
three 30~m telescopes provides rather high detection rate of 10~GeV $\gamma$ rays (see 
Fig.~\ref{fig1}).
 
 \vspace*{-4mm}
 \section{Angular Resolution}
 \vspace*{-2mm}
 
\begin{figure}[t]
\centering
\includegraphics [width=0.44\textwidth]{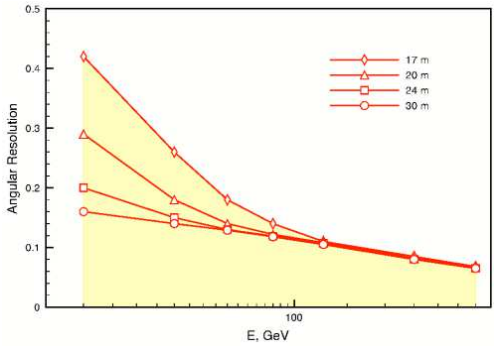}
\vspace*{-4mm}
\caption{Angular resolution ($\sigma$-parameter)  of the array of three telescopes of various aperture sizes.}
\label{fig3}
\vspace*{-3mm}
\end{figure}

For each individual $\gamma$-ray shower the images in all triggered telescopes can be 
used for stereoscopic reconstruction of the shower direction. In the present analysis, only 
`high-quality' images of amplitude grater than 50 ph.-e. with their center of gravity 
located at the angular distance less than $1.2^\circ$ from the camera center are 
accepted for the directional reconstruction. The shower arrival direction, 
i.e. the mean weighted intersection point of major axes in the camera 
focal plane of all available high-quality images, has been computed for each simulated event. Distribution 
of arrival directions for simulated $\gamma$-ray showers (see Fig.~\ref{fig2}) can be fitted to a  
two-dimension Gaussian function. The $\sigma$-parameter of the Gaussian fit stands for the 
angular resolution of the directional reconstruction. Results obtained for telescopes of various aperture 
sizes are summarized in Fig.~\ref{fig3}. Interestingly, an array of three telescopes with larger apertures may noticeably  
improve the angular resolution at low energies (below 100~GeV). The low energy $\gamma$-ray 
showers observed with the telescopes of relatively small aperture generated images of a very small size, 
barely overruning the threshold of 50~ph.-e., and it does not allow an accurate measurement of 
the image orientation. 
In addition, most of these images are located very close to the center of the camera's field of view and 
reveal poorly recognizable elliptical shapes. The telescopes of larger apertures drastically increase the photoelectron content of the recorded images and therefore enable registration of low energy $\gamma$-ray showers at much larger impact distances . 
Both factors help to improve the angular resolution.                            

\vspace*{-4mm}
\section{Background Rejection}
 \vspace*{-2mm}
 
\begin{figure}[t]
\centering
\includegraphics [width=0.44\textwidth]{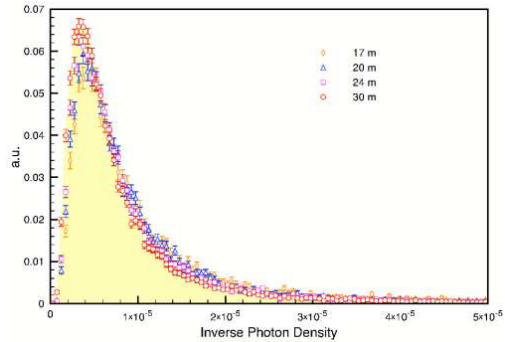}
\vspace*{-4mm}
\caption{Distribution of parameter $\rho^{-1}$ for simulated $\gamma$-ray 
showers of energy from 50 to 100~GeV seen by telescopes of different apertures.}
\label{fig4}
\vspace*{-3mm}
\end{figure}

In fact, the Cherenkov image is a two-dimensional projection of the development of the air shower in space. 
More sensitive telescopes can see more Chrenkov light from the same air shower.  A larger  telescope could 
see Cherenkov emission from many more branches of atmospheric cascades and correspondingly provides 
a better picture of the shower development. A toy model air shower has a spacial shape of a 
spheroid of given width and length.The canonical parameters of image shape, {\it Width} ($w$) 
and {\it Length} ($l$), approximately reproduce the actual spacial extent of the shower. Thus 
the volume of the spheroid containing all charged particles emitting Cherenkov light in the 
shower can be estimated as $V_c \propto w \times w \times l \,\,\, \rm (deg^3)$. The total amount 
of light accumulated in the shower image, {\it Size} ({\it s}), directly relies on the telescope sensitivity 
and primarily its aperture. Such a toy model suggests that the mean spacial density of Cherenkov 
light emitted within the shower volume, $\rho = s/V_c$ (ph.-e./$ \rm deg^3$) is independent of the 
telescope aperture. In Fig.~\ref{fig4} the parameter $\rho^{-1}$ is plotted for $\gamma$-ray  
showers of energy from 50 to 100~GeV registered by the telescopes of various aperture sizes. 
Indeed, the distribution of the 
$\rho^{-1}$-parameter for different telescope apertures show similar shapes. Note that such scaling does not 
work for the cosmic-ray showers, which have irregular spacial profile, and therefore it can be used as an 
additional  selection criterion.   
           
Very high fluctuations in shower development of the low-energy $\gamma$-ray showers is a major limiting 
factor for effective rejection of dominating background cosmic-ray showers. Despite 
the evident advantages in imaging with telescopes of larger apertures,
the rejection power still remains very poor in the 
energy range below 100~GeV, irrespective of telescope aperture. Using an analysis based on standard parameters of image shape ({\it w, l}) the 
rejection factor of cosmic-ray background can be scaled with the shower energy within a 10-100~GeV energy 
range as $q \propto 7 \cdot (E/20\, GeV)^{2/3}$, providing a constant acceptance of $\gamma$-ray showers at the level of about 60\%. Thus, at low energies the rejection power only marginally improves with the rise of the telescope aperture.          

\vspace*{-4mm}
\section{Sensitivity}
\vspace*{-2mm}

Stereoscopic arrays of IACTs of large apertures enable very high detection rates of $\gamma$-ray 
showers of energy from 10 to 100~GeV. It is apparent that the integral 
detection rate of $\gamma$ rays is a strong function of the telescope aperture, $R_\gamma \propto 2 \cdot 
\rm (D/17\,m)^{3/2} (Hz)$ (Crab-like source). The detection area of  20~GeV $\gamma$-ray showers for the 
array of three 30~m telescopes is by a factor of ten larger than that of the array of 17~m telescopes. 
The estimated sensitivity of the array of three 20~m-class telescopes 
is $\rm F_\gamma^{min}(>20\,GeV) \propto 6 \cdot 10^{-11} (D/17\,m)^{-2.5}\, cm^{-2}s^{-1}$. 

\vspace*{-4mm}
\section{Summary}
\vspace*{-2mm}

In the past, the development of instrumentation for the ground-based VHE $\gamma$-rays astronomy 
was mainly driven by the reduction of energy threshold. The low energy 
threshold evidently gives a significant increase in a number of $\gamma$-ray events for the same 
observing time, given the rather steep energy spectra of many well-established TeV $\gamma$-ray 
sources. Expansion into sub-100~GeV energy domain 
brings this technique to unavoidable performance limitations. However, 
future arrays of telescopes of large apertures can still be efficient enough for effective detection of a few tens of GeVs
$\gamma$ rays, and they can be very competitive when measured against other ground-based 
or space-born detectors operating in a similar energy domain.  
    

\bibliographystyle{plain}

\end{document}